\documentclass[english,10pt,aps,pra,twocolumn,groupedaddress,floatfix]{revtex4-1}
 
\usepackage[utf8]{inputenc}
\usepackage{amsmath}
\usepackage{amssymb}
\usepackage{amsfonts}
\usepackage{bm}
\usepackage{bbm}
\usepackage{times}

\usepackage{amsthm}

\usepackage[usenames,dvipsnames]{color}
\usepackage{colortbl}

\definecolor{grey8}{rgb}{.8,.8,.8}
\definecolor{grey9}{rgb}{.9,.9,.9}
\definecolor{dblue}{rgb}{0,0.1,.6}

\usepackage[colorlinks=true,citecolor=dblue,linkcolor=dblue,urlcolor=dblue]{hyperref}

\newcommand{\id}{\mathbbm{1}}
\newcommand{\Tr}{\operatorname{Tr}}
\newcommand{\diag}{\operatorname{diag}}
\renewcommand{\Re}{\operatorname{Re}}
\newcommand{\bra}{\langle}
\newcommand{\ket}{\rangle}
\newcommand{\bbra}{\langle\!\langle}
\newcommand{\kket}{\rangle\!\rangle}
\newcommand{\mc}[1]{\mathcal{#1}}
\newcommand{\pdag}{{\phantom{\dag}}}

\renewcommand{\H}{\mc{H}}
\renewcommand{\L}{\mc{L}}
\newcommand{\D}{\mc{D}}
\newcommand{\M}{\mc{M}}
\newcommand{\N}{\mc{N}}
\newcommand{\G}{\mc{G}}
\newcommand{\B}{\mathfrak{B}}
\newcommand{\s}{\sigma}
\newcommand{\vk}{\vec{k}}
\renewcommand{\vr}{\vec{r}}
\newcommand{\dm}{{\varrho}}
\renewcommand{\vec}[1]{{\boldsymbol{#1}}}

\newcommand{\veps}{\varepsilon}
\newcommand{\tH}{\text{H}}

\newcommand{\ua}{{\uparrow}}
\newcommand{\da}{{\downarrow}}
\newcommand{\ra}{{\rightarrow}}
\newcommand{\la}{{\leftarrow}}
\newcommand{\Ua}{{\Uparrow}}
\newcommand{\Da}{{\Downarrow}}
\newcommand{\Ra}{{\Rightarrow}}

\newcommand{\vac}{\varnothing}

\newtheoremstyle{thmstyle}
  {0.1em} 
  {0em} 
  {\itshape} 
  {} 
  {\bfseries} 
  {.} 
  {.5em} 
  {} 
\theoremstyle{thmstyle} 
\newtheorem{lemma}{Lemma}
\newtheorem{proposition}{Proposition}

\newcommand{\duke} {Department of Physics, Duke University, Durham, North Carolina 27708, USA}

\newcommand{\Title} {Super-operator structures and no-go theorems for dissipative quantum phase transitions}
\newcommand{\Authors}
{
\author{Thomas Barthel}
\affiliation{\duke}
\author{Yikang Zhang}
\affiliation{\duke}
}
\newcommand{\Date} {April 28, 2022}

\begin{document}

\title{\Title\texorpdfstring{\vspace{-0.2em}}{}}
\Authors

\begin{abstract}
In the thermodynamic limit, the steady states of open quantum many-body systems can undergo nonequilibrium phase transitions due to a competition between coherent and driven-dissipative dynamics. Here, we consider Markovian systems and elucidate structures of the Liouville super-operator that generates the time evolution. In many cases of interest, an operator-basis transformation can bring the Liouvillian into a block-triangular form, making it possible to assess its spectrum. The spectral gap sets the asymptotic decay rate. The super-operator structure can be used to bound gaps from below, showing that, in a large class of systems, dissipative phase transitions are actually impossible and that the convergence to steady states follows an exponential temporal decay. Furthermore, when the blocks on the diagonal are Hermitian, the Liouvillian spectra obey Weyl ordering relations. The results apply, for example, to Davies generators and quadratic systems, and are also demonstrated for various spin models.
\end{abstract}

\date{\Date}
\maketitle

\section{Introduction}\label{sec:intro}\vspace{-0.4em}
There are several quantum many-body platforms where a high level of control has been accomplished in the lab.
Prominent examples are circuits of superconducting qubits \cite{Schoelkopf2008-451,Devoret2013-339}, coupled quantum dots (spin qubits) \cite{Loss1998-57,Hanson2007-79}, ultracold atoms in optical lattices or tweezers \cite{Bloch2007,Norcia2018-8,Cooper2018-8} which can also be excited into Rydberg states to achieve strong interactions \cite{Jaksch2000-85,Lukin2001-87,Weimer2010-6}, ions in electromagnetic traps \cite{Cirac1995-74,Blatt2008-453}, and polaritons in circuit-QED or semiconductor-microcavity systems \cite{Hartmann2006-2,Carusotto2013-85}.

Due to practical constraints and our aim of manipulating these systems efficiently, they are inevitably \emph{open} in the sense that they are coupled to the environment. For the trapped atom and ion systems, the employed electronic state space is not entirely isolated. One has, for example, heating of motional degrees of freedom, photoionization, and spontaneous emission, as well as fluctuations in laser beam intensities and pulse timings. Such environment couplings naturally lead to dissipation and decoherence.

As an open system is generally entangled with its environment, it has to be described by a density operator $\dm(t)$. In Markovian systems, the dynamics is then governed by a Lindblad master equation $\partial_t{\dm} = \L(\dm)$ with \cite{Lindblad1976-48,Gorini1976-17,Breuer2007,Rivas2012,Wolf2008-279}
\begin{equation}\label{eq:Lindblad}\textstyle
	\L(\dm)=-i[H,\dm]+\sum_k \gamma_k\Big(L_k^\pdag \dm L_k^\dag-\frac{1}{2} \{L_k^\dag L_k^\pdag,\dm\}\Big).
\end{equation}
In addition to the unitary Hamiltonian part, the Liouville super-operator $\L$ contains dissipative terms, where Lindblad operators $L_k$ describe the environment couplings with strengths $\gamma_k\geq 0$. At long times, systems with a finite-dimensional Hilbert space converge to \emph{steady states} $\dm_\text{ss}$ obeying $\L(\dm_\text{ss})=0$ or evolve unitarily in an asymptotic subspace \cite{Baumgartner2008-41}. All eigenvalues of $\L$ have non-positive real parts \cite{Rivas2012}, and the largest nonzero eigenvalue real part sets the asymptotic decay rate -- the so-called Liouville gap $\Delta$.
The exceptional control in the aforementioned experimental systems can be used to design and tune the environment couplings. In this way, one could stabilize useful states \cite{Kraus2008-78,Diehl2008-4,Verstraete2009-5} such as cluster or more general graph states \cite{Briegel2001-86,Hein2004-69} for measurement-based quantum computation \cite{Raussendorf2001-86,Walther2005-434}, or suitable initial states for quantum phase estimation \cite{Abrams1999-83,AspuruGuzik2005-309} and quantum simulation. Moreover, in a competition of coherent and driven-dissipative dynamics, one could drive the system into (novel) phases of matter that may not be accessible in equilibrium.

Here, we investigate conditions under which phase transition in the steady state can or cannot occur on a fundamental level.
In the literature, the theoretical analysis is often based on mean-field approximations: For many-body lattice systems, interaction terms are decoupled in the mean field theory to arrive at nonlinear equations of motion for local observables. The mean-field solutions may suggest rich phase diagrams and can include multi-stable and oscillatory phases \cite{Szymanska2007-75,Diehl2008-4,Diehl2010-105b,Tomadin2011-83,Tomadin2012-86,Lee2011-84,Lee2012-108,Lee2013-110,Boite2013-110,Boite2014-90,Marcuzzi2014-113,Naether2015-91,Wilson2016-94,Biondi2017-96,Savona2017-96}. The validity of such results is often unclear as mean field theory is notoriously unreliable for dynamical problems, even for closed systems. In several cases, qualitative discrepancies have been found in comparison with other approaches \cite{Lee2013-110,Jin2013-110,Hoening2014-90,Weimer2015-114,Weimer2015-91,Maghrebi2016-93,MendozaArenas2016-93,FossFeig2017-95,Vicentini2018-97,Huber2020-102,Rota2019-122}.

Like quantum phase transitions in closed systems \cite{Sachdev2011}, a dissipative phase transition is characterized by a non-analytic dependence of steady-state expectation values on system parameters. This, in turn, requires a non-analytic change in the steady-state density matrix and, hence, a level crossing \cite{Kato1995}. So, the spectral gap $\Delta$ of the Liouvillian needs to close at the transition point \cite{Kessler2012-86,Minganti2018-98}.

In this paper, we describe classes of Liouvillians that can be brought into a block-triangular form by operator-basis transformations. These transformations are in general non-orthogonal and can, for example, be found using symmetries and other dynamical constraints. The Liouvillian spectrum can then be assessed through the spectra of the blocks on the diagonal. If the blocks are all Hermitian, the spectrum is completely real. The eigenvalues of sums of such Liouvillians obey Weyl ordering relations, which can result in bounds on the gap and, hence, preclude steady-state phase transitions (Sec.~\ref{sec:Real}). This class of open systems encompasses, for example, the Davies generators \cite{Davies1979-34,Spohn1978-19,Spohn1978-38,Alicki2007} that arise in the context of thermalization for systems that are weakly coupled to thermal baths. We then generalize further to Liouvillians that can be transformed to block-triangular form but have non-Hermitian blocks on the diagonal, and we discuss ways to bound their spectral gaps (Secs.~\ref{sec:BT-nH}, \ref{sec:BT-nH-example}, and \ref{sec:BT-nH-quadratic}). All results are illustrated with exemplary many-body lattice models. The block-triangular structures can make (seemingly) perturbative treatments exact in the sense that effective Liouvillians coincide with specific diagonal blocks in the block-triangular representation and, hence, yield exact Liouvillian eigenvalues. Exploiting the block structure also makes numerical approaches more efficient and can lead to integrable subproblems.

\section{A dissipative model with \texorpdfstring{$\mathbb{Z}_2$}{Z2} symmetry breaking}\label{sec:Z2model}
Let us start the discussion with a concrete spin-1/2 system on a cubic lattice of $N$ sites in $d$ dimensions. It serves as a motivation and example for the general discussion in Sec.~\ref{sec:Real}. We can try to mimic the physics of the non-dissipative transverse Ising model at zero temperature. The paramagnetic state $|\Ra\ket:=|\ra\ra\dots\ra\ket$ with $\s_i^x|\Ra\ket=|\Ra\ket$ can be stabilized by Lindblad operators $L^x_i:=\s_i^{x+}$. The ferromagnetic states $|\Ua\ket:=|\ua\ua\dots\ua\ket$ and $|\Da\ket$ are stabilized by controlled spin-flip operators
\begin{equation*}
	L^+_{ij}:=P^\ua_i\s_j^++\s_i^+P^\ua_j\quad\text{and}\quad
	L^-_{ij}:=P^\da_i\s_j^-+\s_i^-P^\da_j,
\end{equation*}
acting on bonds $(i,j)$. Here, $\s^{\alpha}_i$ are the Pauli matrices on site $i$, $\s^{x+}=|\ra\ket\bra\la|$, and $P^s_i$ projects onto $\s^z$ or $\s^x$ eigenstates $|s\ket$ on site $i$ with $s\in\{\ua,\da,\ra,\la\}$. For simplicity, we include no Hamiltonian terms such that
\begin{subequations}\label{eq:openTransvIsing}
\begin{align}\label{eq:openTransvIsing-Full}
	&\L=\gamma_x \D^x+\gamma_f\underbrace{(\D^++\D^-)}_{=:\D^f}+\gamma_z\D^z\quad\text{with}\\\label{eq:openTransvIsing-x}
	&\D^x(\dm)=\sum_i\Big(\s_i^{x+}\dm\s_i^{x-}-\frac{1}{2}\{P^\la_i,\dm\}\Big),\\\label{eq:openTransvIsing-pm}
	&\D^\pm(\dm)=\sum_{\bra i,j\ket}\Big(L^{\pm\pdag}_{ij}\dm (L^{\pm}_{ij})^\dag-\frac{1}{2}\{Q_{ij},\dm\}\Big),\\
	&\text{and}\quad \D^z(\dm)=\sum_i\left(\s_i^{z} \dm\s_i^{z}-\dm\right).
\end{align}
\end{subequations}
The sum in Eq.~\eqref{eq:openTransvIsing-pm} runs over all lattice bonds, and $Q_{ij}=(L^{\pm}_{ij})^\dag L^{\pm\pdag}_{ij}=\frac{1}{2}(\id-\s^z_i\s^z_j)+\s^+_i\s^-_j+\s^-_i\s^+_j$. The term $\D^z$ has been included to make $|\Ua\ket\bra\Ua|$ and $|\Da\ket\bra\Da|$ the only steady states for $\gamma_x\to 0$. The system has a $\mathbb{Z}_2$ symmetry generated by $U^x=\prod_i\s^x_i$.

\subsection{The disordered phase}\label{sec:Z2-disordered}
For $\gamma_x\gg\gamma_f,\gamma_z$, the system is in the disordered (para\-mag\-net\-ic) phase. At $\gamma_f=\gamma_z=0$, the non-degenerate steady state is $|\Ra\ket\bra\Ra|=\prod_i P^\ra_i$. The first excitations can be generated by flipping one spin in the bra or ket and have eigenvalue $\lambda=-\gamma_x/2$. Noting that $\sigma^y$ and $\sigma^z$ are linear combinations of $|\la\ket\bra\ra|$ and $|\ra\ket\bra\la|$, we choose the operators
\begin{equation}
	A_i^z:=\sigma^z_i\prod_{j\neq i}P^\ra_j\quad\text{and}\quad
	A_i^y:=\sigma^y_i\prod_{j\neq i}P^\ra_j
\end{equation}
as a basis for the subspace of first excitations such that $\L(A^\alpha_i)=-\frac{\gamma_x}{2} A^\alpha_i$. The corresponding left eigenvectors of $\L$ are $B_i^z=\sigma^z_i/2$ and $B_i^y=\sigma^y_i/2$. They obey the biorthonormality conditions
\begin{equation}\label{eq:biortho}
	\bbra B_i^\beta|A_j^{\alpha}\kket\equiv \Tr\big( B_i^{\beta\dag} A_j^{\alpha}\big)
	=\delta_{i,j}\delta_{\alpha,\beta}
\end{equation}
Taking $\D^\pm$ and $\D^z$ into account perturbatively, the extensive degeneracy of the first excitations is lifted. For a $d$-dimensional cubic lattice, the only nonzero matrix elements of the effective Liouvillian $\bbra B^\alpha_i|\L|A^\beta_j\kket\equiv \bbra B^\alpha_i|\L(A^\beta_j)\kket$ are
\begin{subequations}\label{eq:Z2-Leff}
\begin{align}
	&\bbra B^z_i|\L|A^z_i\kket=-\gamma_x/2-2d\gamma_f\quad\forall i,\\
	&\bbra B^y_i|\L|A^y_i\kket=-\gamma_x/2-2d\gamma_f-2\gamma_z\quad\forall i,\\
	&\bbra B^z_i|\L|A^z_j\kket=\gamma_f,\ \  \bbra B^y_i|\L|A^y_j\kket=-\gamma_f\quad\forall \bra i,j\ket.
\end{align}
\end{subequations}
So, with site positions $\vr_j\in\mathbb{Z}^d$ and $\alpha\in\{z,y\}$, the excitations become $A^\alpha_{\vk}=\frac{1}{N}\sum_j e^{i\vk\cdot\vr_j}A^\alpha_j$ with crystal momentum $\vk$ and dispersions
\begin{subequations}\label{eq:Z2-dispersions}
\begin{align}
	&\lambda^z_\vk=-\gamma_x/2-2\gamma_f\sum_{a=1}^d(1-\cos k_a),\\
	&\lambda^y_\vk=-\gamma_x/2-2\gamma_f\sum_{a=1}^d(1+\cos k_a)-2\gamma_z
\end{align}
\end{subequations}
up to higher-order corrections. Thus, to leading order, the Liouvillian gap stays fixed with $-\lambda^z_{\vk=\vec{0}}=\gamma_x/2$.

\subsection{The ferromagnetic point}\label{sec:Z2-ferro}
At the ferromagnetic point $\gamma_x=0$, the $\mathbb{Z}_2$ symmetry is spontaneously broken with steady states $|\Ua\ket\bra\Ua|$ and $|\Da\ket\bra\Da|$. Without the dissipator $\D^z$, projections $|\Phi\ket\bra\Phi|$ onto superpositions $|\Phi\ket=\alpha|\Ua\ket+\beta|\Da\ket$ would also be steady states. For convenience, let us use the linear combinations
\begin{equation}
	\dm_\pm:=\frac{1}{2}\left(|\Ua\ket\bra\Ua|\pm|\Da\ket\bra\Da|\right).
\end{equation}
The corresponding left zero-eigenvalue eigenvectors of $\gamma_f\D^f+\gamma_z\D^z$ are
\begin{equation}\textstyle
	B_+=\id\quad\text{and}\quad B_-=\frac{1}{N}\sum_i\s_i^z.
\end{equation}
They represent conserved quantities and obey biorthonormality, $\bbra B_\veps |\dm_{\veps'}\kket=\delta_{\veps,\veps'}$. To take $\D^x$ into account perturbatively, one finds $\bbra B_\veps|\D^x|\dm_{\veps'}\kket=-\delta_{\veps,\veps'}\delta_{\veps,-}/2$. Hence, to leading order in $\gamma_x$, $\dm_+$ becomes the unique steady state ($\lambda_+=0$), and the degeneracy is immediately broken with eigenvalue $\lambda_-=-\gamma_x/2$ for the first excitation $\dm_-$. So, the $\mathbb{Z}_2$ symmetry is recovered for $\gamma_x>0$, and we obtain the same gap as in the disordered phase.

A continuum of low-lying excitations at $\gamma_x=0$ consists of single-sided magnons, i.e., operators $A^\alpha_\vk=\frac{1}{N}\sum_j e^{i\vk\cdot\vr_j}A^\alpha_j$ with $A^{1}_j=\sigma^x_j|\Ua\ket\bra\Ua|$, $A^{2}_j=|\Ua\ket\bra\Ua|\sigma^x_j$, $A^{3}_j=\sigma^x_j|\Da\ket\bra\Da|$, and $A^{4}_j=|\Da\ket\bra\Da|\sigma^x_j$. They have the four-fold degenerate eigenvalues $\lambda_\vk=-2\gamma_z-2\gamma_f\sum_{a=1}^d(1+\cos k_a)$. The degeneracy is partially lifted in first-order perturbation theory, shifting the eigenvalue $\lambda_\vk$ by $-\gamma_x/2$, $-\gamma_x$ (twice), or $-3\gamma_x/2$.

Numerically, one also finds an isolated eigenvalue $\lambda_b$ in-between the first excitation and the single-magnon continuum. For small $\gamma_x$, it obeys
$-\gamma_x/2>\lambda_b>-2\gamma_z-\gamma_x/2$. The nature of this excitation will be clarified in Sec.~\ref{sec:Real-Z2model}.

\subsection{Further observations and interpretation}
Beyond perturbation theory, it turns out that all eigenvalues of the Liouvillian \eqref{eq:openTransvIsing} are \emph{real} and that, surprisingly, the gap is fixed at $\gamma_x/2$ in the \emph{entire phase diagram}. At this point, it is evident that the analogy to the Hamiltonian transverse Ising model does not carry very far here: We have an isolated transition point at $\gamma_x=0$ without a continuum of gapless excitations. There is $\mathbb{Z}_2$ symmetry breaking at this point, but we do not actually have an extended ferromagnetic phase. Also, the excitations above the ferromagnetic steady states are magnons instead of domain walls. While this can be rectified in a modified model, the gap being fixed at $\gamma_x/2$ in this system is remarkable and lead to an interesting general finding: 

The unapparent cause for the observed spectral properties is that one can apply a non-orthogonal basis transformation such that $\D^x$, $\D^\pm$, and $\D^z$ assume a block-triangular form with Hermitian blocks on the diagonal. This has far-reaching consequences, as discussed in the following.

\section{Liouvillians with real spectrum}\label{sec:Real}
The reason for the gap in the model \eqref{eq:openTransvIsing} being finite for all nonzero $\gamma_x$ and further properties of its Liouvillian spectrum are closely related to the fact that the eigenvalues of the competing terms $\gamma_x\D^x$ and $\gamma_f \D^f+\gamma_z\D^z$ are all real. Let us discuss, more generally, conditions under which Liouvillians have an entirely real spectrum.

\subsection{Hermiticity}
The simplest scenario are Liouvillians that are self-adjoint $\L=\L^\dag$ and, hence, have a real spectrum. Here the adjoint of a super-operator $\M$ is defined through the Hilbert-Schmidt inner product in Eq.~\eqref{eq:biortho} such that $\bbra B|\M(A)\kket=\bbra\M^\dag(B)|A\kket$. In particular, the adjoint
\begin{equation}\label{eq:LindbladAdj}
	\L^\dag(B)=i[H,B]+\sum_k \gamma_k\Big(L_k^\dag B L_k^\pdag-\frac{1}{2} \{L_k^\dag L_k^\pdag,B\}\Big)
\end{equation}
of the Liouvillian \eqref{eq:Lindblad} generates the time evolution in the Heisenberg picture, $\partial_t B=\L^\dag(B)$. To construct a self-adjoint Liouvillian, one can for example choose $H=0$ and dissipators such that for every Lindblad operator $L_k$ with rate $\gamma_k$, there exists a second term with $\sqrt{\gamma_{k'}}L_{k'}=\sqrt{\gamma_{k}}L_{k}^\dag$.

Recalling that the spectrum of a matrix is invariant under similarity transformations, a more general class of models is captured by the following.

\begin{lemma}\label{sttmnt:Real-selfadj}
  $\L$ has a real spectrum if we can find an (in general, non-orthogonal) basis in which it is Hermitian, i.e., if there exists an invertible super-operator $\G$ such that $\G\L\G^{-1}=(\G\L\G^{-1})^\dag$.
\end{lemma}

A simple example is the dissipator $\D^x=\sum_i\D_i^x$ for the Lindblad operators $\s_i^{x+}$ in [Eq.~\eqref{eq:openTransvIsing-x}]. For the orthonormal local operator basis $(|\ra\ket\bra\ra|,|\la\ket\bra\la|,|\la\ket\bra\ra|,|\ra\ket\bra\la|)$, it is non-Hermitian; hence, it is non-Hermitian in all orthonormal bases. In particular, it has the off-diagonal element $\D_i^x(|\la\ket\bra\la|)=|\ra\ket\bra\ra|$, but $\D_i^x(|\ra\ket\bra\ra|)=0$. In contrast, choosing the non-orthogonal right basis
\begin{equation}\label{eq:Bx1}
	\B_x:=(|\ra\ket\bra\ra|,\s^x,\s^z,\s^y),
\end{equation}
it attains the diagonal form 
\begin{equation}\label{eq:Dx-in-Bx}
	[\D_i^x]_{\B_x}=\diag(0,-1,-1/2,-1/2),
\end{equation}
and is hence Hermitian. Here, $[\D_i^x]_{\B_x}$ denotes the matrix with elements $\bbra B_m|\D_i^x|A_n\kket$ for $A_n\in \B_x$ and corresponding left basis elements $B_m\in(\id,-|\la\ket\bra\la|,\s^z/2,\s^y/2)$ that obey biorthonormality $\bbra B_m|A_n\kket=\delta_{m,n}$.

A well-known class of Liouvillians, covered by Lemma~\ref{sttmnt:Real-selfadj}, are \emph{Davies generators} \cite{Davies1979-34,Spohn1978-19,Spohn1978-38,Alicki2007,Roga2010-66,Temme2013-54,Shabani2016-94,Kastoryano2016-344,Wilming2017-58,Metcalf2020-2}. They arise when a system is weakly coupled to the environment through Hermitian system operators $S_k$ and bath operators, and when the bath is in thermal equilibrium. The Lindblad operators $L_k(\omega)$ are then given by the Fourier coefficients of $S_k$, i.e., $e^{iHt}S_ke^{-iHt}=:\sum_\omega e^{i\omega t}L_k(\omega)$, where the sum runs over all eigenenergy differences $\omega=E_n-E_m$. The corresponding rates $\gamma_k(\omega)$ in Eq.~\eqref{eq:Lindblad} obey the Kubo-Martin-Schwinger (KMS) condition $\gamma_k(-\omega)=e^{-\beta H}\gamma_k(\omega)$ with the temperature $\beta^{-1}$ being set by the environment state. As $L_k(\omega)$ connects $H$ eigenstates with energy difference $\omega$, it obeys $e^{-\beta H}L_k(\omega)=e^{\beta\omega}L_k(\omega)e^{-\beta H}$. It follows that Davies generators can describe thermalization in the sense that the canonical ensemble $\dm_\beta=e^{-\beta H}/Z$ is a steady state and that the so-called detailed-balance condition
\begin{equation}
	\G_\beta \L=\L^\dag \G_\beta \quad\text{with}\quad
	\G_\beta(X):=\dm_\beta^{1/2} X \dm_\beta^{1/2}
\end{equation}
is obeyed. This shows that Lemma~\ref{sttmnt:Real-selfadj} applies with $\G=\G_\beta^{1/2}$. Note that our previous example, $\D^x$, is \emph{not} a Davies generator; for example, its steady state does not have full rank.

\subsection{Block-triangular with Hermiticity}
In order to understand the observations made for the spin model in Sec.~\ref{sec:Z2model}, we need to generalize further.

\begin{proposition}\label{sttmnt:Real-BTH}
  If we can find an (in general, non-orthogonal) operator basis in which $\L$ is block-triangular with Hermitian blocks $\M_i$ on the diagonal (BTH), i.e., if there exists an invertible super-operator $\G$ such that 
 \begin{equation}\label{eq:BTH}
  \arraycolsep=0pt\def\arraystretch{1.2}
  \G\L \G^{-1}=\left\lgroup
  \begin{array}{ccc}
     \cellcolor{grey8} \M_1 &                        &  \\
     \cellcolor{grey9}      & \cellcolor{grey8} \M_2 &  \\
     \cellcolor{grey9}      & \cellcolor{grey9}      & \cellcolor{grey8} \pdag\ddots\pdag
  \end{array}
  \right\rgroup \,\,\text{or}\,\,
  \left\lgroup
  \begin{array}{ccc}
     \cellcolor{grey8} \M_1 & \cellcolor{grey9}      & \cellcolor{grey9} \\
                            & \cellcolor{grey8} \M_2 & \cellcolor{grey9} \\
                            &                        & \cellcolor{grey8} \pdag\ddots\pdag
  \end{array}
  \right\rgroup
 \end{equation}
 with $\M_i^\pdag=\M_i^\dag$, then the spectrum of $\L$ is real, consisting of the eigenvalues of the matrices $\M_i$.
\end{proposition}

This is due to the fact that the spectrum of a triangular matrix is given by its diagonal elements and, in Eq.~\eqref{eq:BTH}, $\G\L \G^{-1}$ is transformed to lower or upper triangular form when diagonalizing the blocks $\M_i$. For the typical many-body systems that we consider, it is of course practically impossible to transform the Liouvillian to Jordan normal form (or to diagonalize all blocks $\M_i$). As we will see in examples, the idea behind Proposition~1 is to use symmetries or other properties of the system to attain a BTH form \eqref{eq:BTH}.

As a simple example we can again invoke the dissipator $\D_i^x$ [Eq.~\eqref{eq:openTransvIsing-x}]. For a single site, it is upper triangular in the basis $(|\ra\ket\bra\ra|,|\la\ket\bra\la|,|\la\ket\bra\ra|,|\ra\ket\bra\la|)$ with the diagonal elements $(0,-1,-1/2,-1/2)$ giving the real spectrum. Hence, the full dissipator $\D^x=\sum_i\D^x_i$ is also upper triangular when using the corresponding tensor product basis.

A more intricate example is the dissipator $\D^f=\D^++\D^-$ of the spin system in Eq.~\eqref{eq:openTransvIsing}. $\D^f$ can be brought into a lower block-triangular form when using $\B_x$ [Eq.~\eqref{eq:Bx1}] as the single-site operator basis. And if we modify the basis slightly to
\begin{equation}\label{eq:Bx2}
	\B'_x:=(|\ra\ket\bra\ra|,\s^x,2^{1/4}\s^z,\s^y),
\end{equation}
all blocks on the diagonal become Hermitian, such that $[\D^f]_{\B'_x}$ is BTH, showing that its spectrum is real. The nonzero matrix elements are given in Table~\ref{tab:Df-Bx2}. The global BTH structure of the many-body system will be discussed in more detail in Sec.~\ref{sec:Real-Z2model}.
\begin{table}[t]
	\setlength{\tabcolsep}{3ex}
	\begin{tabular}{c  l}
	 \hline\\[-0.8em]
	 $\dm$ & $\D^f(\dm)$\\[0.2em]
	 \hline\\[-0.8em]
	 $\vac\vac$	& $-\vac x-x\vac+\frac{1}{2}xx+\frac{1}{\sqrt{2}}zz+\frac{1}{2}yy$\\[0.2em]
	 $\vac x$	& $-\vac x-x\vac+\frac{1}{2}xx+\frac{1}{\sqrt{2}}zz+\frac{1}{2}yy$\\
	 $\vac z$	& $-\vac z+z\vac-zx$\\
	 $\vac y$	& $-\vac y-y\vac+yx$\\[0.2em]
	 $xx$	& $\,\,\,\,-xx+\sqrt{2}zz+\,\,\quad yy$\\
	 $zz$	& $\sqrt{2}xx-\,\,\,\,2\,zz-\sqrt{2}yy$\\
	 $yy$	& $\,\,\quad xx-\sqrt{2}zz-\,\,\quad yy$\\[0.2em]
	 $xz$	& $-xz-zx$\\
	 $xy$	& $-xy+yx$\\
	 $zy$	& $-zy-yz$\\[0.2em]
	\hline
	\end{tabular}
	\caption{\label{tab:Df-Bx2}Nonzero matrix elements of the reflection-symmetric dissipator $\D^f=\D^++\D^-$ in Eq.~\eqref{eq:openTransvIsing-Full} on a single bond with respect to the operator basis $\B_x'$ on both sites. The basis elements, given in Eq.~\eqref{eq:Bx2}, are here denoted by $\vac$, $x$, $z$, and $y$.}
\end{table}

\subsection{Weyl ordering and a no-go theorem for phase transitions}
Let the Liouvillians $\L=\L^{(1)}+\L^{(2)}$ consist of two terms that are BTH with the same block structure and have descendingly ordered eigenvalues $0=\lambda^{(i)}_1\geq \lambda^{(i)}_2\geq\dots$ with $i=1,2$. Then, the eigenvalues $0=\lambda_1\geq\lambda_2\geq\dots$ of $\L$ are also real and obey the ordering relations
\begin{equation}\label{eq:WeylOrder}
	\lambda_k\leq \lambda_k^{(i)}\quad\text{and}\quad
	\lambda_k\geq \lambda_k^{(1)}+\min_q\lambda_q^{(2)}\quad\forall\, k.
\end{equation}
This follows by application of Weyl's theorem \cite{Bhatia1997} to the blocks on the diagonal. In particular, if we have two Hermitian $n\times n$ matrices $\M$ and $\N$, and denote their descendingly ordered eigenvalues by $\lambda_k(\M)$ and $\lambda_k(\N)$, then one can use the variational characterization of eigenvalues according to the minmax principle \cite{Bhatia1997} to show that $\lambda_k(\M)+\lambda_n(\N)\leq \lambda_k(\M+\N)\leq \lambda_k(\M)+\lambda_1(\N)$ for all $k\in[1,n]$. For Liouvillians with a real spectrum, the largest eigenvalue is necessarily $\lambda_1=0$, corresponding to a steady state, and we thus obtain Eq.~\eqref{eq:WeylOrder}.
With this, we arrive at the following no-go theorem for dissipative phase transitions.

\begin{proposition}\label{sttmnt:Real-NoGo}
  Consider $\L(\vec{\gamma})=\gamma_0\L^{(0)}+\sum_\nu\gamma_\nu\L^{(\nu)}$ with a basis in which the Liouvillians $\L^{(k)}$ are all BTH with the same block structure. The spectra of $\L(\vec{\gamma})$ and its components then obey the Weyl ordering relations \eqref{eq:WeylOrder}. Specifically, if $\L^{(0)}$ has a unique steady state and finite gap $\Delta$, then the gap of $\L(\vec{\gamma})$ is at least $\gamma_0\Delta$, and the system can only be critical for $\gamma_0=0$.
\end{proposition}

As discussed in the following section, this result explains, for example, why the gap in the spin-1/2 model \eqref{eq:openTransvIsing} cannot be smaller than $\gamma_x/2$ in the entire phase diagram.

\subsection{Application to the dissipative model of Sec.~\ref{sec:Z2model}}\label{sec:Real-Z2model}
Let us revisit the spin-1/2 model $\L=\gamma_x \D^x+\gamma_f\D^f+\gamma_z\D^z$ [Eq.~\eqref{eq:openTransvIsing}], which features $\mathbb{Z}_2$ symmetry breaking at the ferromagnetic point $\gamma_x=0$, and let us explain its properties in the light of Proposition~\ref{sttmnt:Real-NoGo}. In the non-orthogonal single-site basis \eqref{eq:Bx2}, the dissipators $\D^x=\sum_i\D_i^x$ and $\D^z=\sum_i\D_i^z$ become diagonal and lower triangular, respectively, where $[\D_i^x]_{\B_x'}=\diag(0,-1,-1/2,-1/2)$ and
\begin{equation}
	[\D_i^z]_{\B_x'}=\begin{pmatrix}0&0&0&0\\-1&-2&0&0\\0&0&0&0\\0&0&0&-2\end{pmatrix}.
\end{equation}
Also, $\D^f$ assumes lower BTH form with the nonzero elements given in Table~\ref{tab:Df-Bx2}.

With the dissipator $\D^x$ in mind, let us refer to the elements $(|\ra\ket\bra\ra|,\s^x,2^{1/4}\s^z,\s^y)$ of the operator basis \eqref{eq:Bx2} as the \emph{vacuum} $\vac$, and $x$, $z$, and $y$ \emph{particles}, respectively. We can then characterize the actions of the three dissipators in that basis as follows. $\D^x$ is diagonal. $\D^z$ is diagonal except for terms that generate a $z$ particle from the vacuum ($\vac\mapsto z$). The non-diagonal elements of the reflection-symmetric dissipator $\D^f$ can create one $x$ particle or a particle pair of type $x$, $y$, or $z$ from the vacuum ($\vac\vac\mapsto \vac x,xx,yy,zz$). For a bond with an $x$ particle and one vacuum site, $\D^f$ can let the particle hop or create a particle pair ($\vac x\mapsto x\vac,xx,yy,zz$). When we have a bond with a $y$ or $z$ particle and one vacuum site, $\D^f$ can let the particle hop or create an $x$ particle ($\vac y\mapsto y\vac,yx$; $\vac z\mapsto z\vac,zx$). When two particles of equal flavor meet on a bond, $\D^f$ can transform them into a pair of the other two flavors ($xx\leftrightarrow yy\leftrightarrow zz\leftrightarrow xx$). When two particles of different flavor meet on a bond, $\D^f$ can swap their positions ($\alpha\beta\leftrightarrow \beta\alpha$).

Thus, the total number of particles never decreases under the action of $\L$, and the particle number parities $P_\alpha=N_\alpha \bmod 2$ are conserved. So, we can structure the operator space into blocks, ordered by increasing particle number $N_x+N_y+N_z$ and can further decompose into sectors with different parities $(P_x,P_y,P_z)$. All blocks on the diagonal are Hermitian, and $\L$ is lower block-triangular.

The dissipator $\D^x$ has the spectral gap $1/2$ and, according to Proposition~\ref{sttmnt:Real-NoGo}, the gap of $\L$ can hence never fall below $\gamma_x/2$. This excludes steady-state phase transitions at nonzero $\gamma_x$. The perturbative treatment in Sec.~\ref{sec:Z2-disordered} suggested that the gap remains fixed at $\gamma_x/2$ for all $\gamma_f$ and $\gamma_z$; it was associated with the $\vk=\vec{0}$ momentum state of a single $z$ particle. According to the block-triangular structure described above, the perturbatively determined eigenvalues \eqref{eq:Z2-dispersions} of $\L$ are actually \emph{exact}: The effective Liouvillian $\bbra B^\alpha_i|\L|A^\beta_j\kket$ in Eq.~\eqref{eq:Z2-Leff} simply consists of the single $y$- and $z$-particle blocks on the diagonal in our chosen basis $\B_x'$. As pointed out in Proposition~\ref{sttmnt:Real-BTH}, their eigenvalues are exact $\L$ eigenvalues. (This is, of course, not true for the corresponding eigenvectors.) The largest block eigenvalue $-\gamma_x/2$ is an exact eigenvalue of $\L$ for all $\gamma_f$ and $\gamma_z$. Finally, we can convince ourselves that it is always the overall largest nonzero $\L$ eigenvalue: The single $x$-particle block on the diagonal also features a cosine dispersion with the largest eigenvalue $-\gamma_x-2\gamma_z$ corresponding to a state with momentum $k_a=\pi$ for $a=1,\dotsc,d$. For the blocks with total particle number $N>1$, an upper spectral bound of $-N\gamma_x/2$ follows from the diagonal elements due to $\D^x$. Thus, the spectral gap of $\L$ is indeed exactly $\gamma_x/2$ for all $\gamma_x,\gamma_f,\gamma_z$.

Finally, let us comment on the isolated second excitation with eigenvalue $\lambda_b$ above the single-magnon continuum found in the perturbative analysis for small $\gamma_x$ in Sec.~\ref{sec:Z2-ferro}. It turns out to be the largest eigenvalue of the two-particle block on the diagonal with even parities $P_x=P_y=P_z=0$ and corresponds to a two-particle bound state.

\section{Bounding gaps of Liouvillians with non-Hermitian blocks}\label{sec:BT-nH}
Propositions~\ref{sttmnt:Real-BTH} and \ref{sttmnt:Real-NoGo} cover BTH Liouvillians. More generally, we now address Liouvillians that are block-triangular in a suitable basis but may have blocks $\M_i$ on the diagonal that are neither Hermitian nor anti-Hermitian. As an example, in any orthonormal operator basis, Hamiltonian terms $\H(\dm):=-i[H,\dm]$ are anti-Hermitian because $\H^\dag(B)=i[H,B]=-\H(B)$.
For the study of spectral gaps, it would of course be sufficient for the block that contains the first excitation to be Hermitian. If this form is difficult or impossible to attain, other options are to study the singular values of $\M_i$ or to bound the gap. The latter can, e.g., be done using the eigenvalues of the Hermitian components
\begin{equation}\label{eq:MH}
	\M_i^\tH:=(\M_i^\pdag+\M_i^\dag)/2
\end{equation}
or Gershgorin's circle theorem, as discussed in the following.

\begin{proposition}\label{sttmnt:BT-nH}
  If we can find a (basis) transformation $\G$ such that $\L$ assumes the block-triangular form
  \begin{equation*}
  \arraycolsep=0pt\def\arraystretch{1.2}
  \G\L \G^{-1}=\left\lgroup
  \begin{array}{cccc}
     \cellcolor{grey8} \pdag0\pdag &                        &                        &  \\
     \cellcolor{grey9}             & \cellcolor{grey8} \M_1 &                        &  \\
     \cellcolor{grey9}             & \cellcolor{grey9}      & \cellcolor{grey8} \M_2 &  \\
     \cellcolor{grey9}             & \cellcolor{grey9}      & \cellcolor{grey9}      & \cellcolor{grey8} \pdag\ddots\pdag
  \end{array}
  \right\rgroup \,\,\text{or}\,\,
  \left\lgroup
  \begin{array}{cccc}
     \cellcolor{grey8} \pdag0\pdag & \cellcolor{grey9}      & \cellcolor{grey9}      & \cellcolor{grey9} \\
                                   & \cellcolor{grey8} \M_1 & \cellcolor{grey9}      & \cellcolor{grey9} \\
                                   &                        & \cellcolor{grey8} \M_2 & \cellcolor{grey9} \\
                                   &                        &                        & \cellcolor{grey8} \pdag\ddots\pdag
  \end{array}
  \right\rgroup,
 \end{equation*}
 and if we can find negative upper bounds on the largest eigenvalue real parts of the block matrices $\M_i^\pdag$ ($\neq\M_i^\dag$), then no phase transitions can occur.
\end{proposition}

Again, the idea is to exploit symmetries or other properties of $\L$ to attain such a block-triangular form. The cost for transforming into Jordan normal form generally grows exponentially in the system size.
To achieve that the zero eigenvalue is located in the first $1\times 1$ block of $\G\L \G^{-1}$, the first right basis element should be chosen as the actual steady state or, if it is not known, at least the first left basis element should be the identity. In these cases, the spectral gap $\Delta$ of $\L$ is determined by the largest real part of the eigenvalues of the blocks $\M_{i>0}$, i.e.,
\begin{equation}
	\Delta=-\max_{i>0,k}\Re\lambda_k(\M_i).
\end{equation}
If the asymptotic subspace ($\Re \lambda_k=0$) cannot be separated in the block structure from further excitations, one needs to additionally assess the largest nonzero eigenvalue real part of any block that contains an element of the asymptotic subspace. While this generalization is straightforward, for conciseness, we assume the former scenario in the following.

\emph{Using block singular values.} --
If the $\M_i$ eigenvalue $\lambda$ with largest real part is real, it can be determined by the smallest singular value $\nu$ of $\M_i$ such that
\begin{equation}
	\max_k\Re\lambda_k(\M_i)=-\nu.
\end{equation}
It can also be obtained variationally as
\begin{equation}\label{eq:Bound-SVD-var}
	\nu^2=-\inf_{\|A\|=1}\bbra A|\M_i^\dag\M_i^\pdag|A\kket
\end{equation}
Let $\lambda=-\Delta+i\varphi$ be the $\M_i$ eigenvalue with largest real part. If the imaginary part $\varphi$ is nonzero, the corresponding singular value $\nu=\sqrt{\Delta^2+\varphi^2}$ would only provide an \emph{upper bound} on $\Delta$. But if it can be established that $\varphi=0$, then $\nu=\Delta$. In general, it is unfortunately difficult to assess whether $\lambda$ is real. Even if the smallest singular value is unique, we only know that it corresponds to \emph{a} real eigenvalue, and, in general, it only gives an upper bound on $\Delta$.

\emph{Using the Hermitian component.} --
A more straightforward approach is to study the spectrum of the Hermitian component \eqref{eq:MH}.
The largest and smallest eigenvalues $\mu_{\max}$ and $\mu_{\min}$ of the Hermitian component $\M_i^\tH$ of a Liouvillian block $\M_i$ bound the real parts of its eigenvalues such that
\begin{equation}\label{eq:Bound-H}
	\mu_{\max}\geq \Re\lambda_k(\M_i)\geq \mu_{\min}\quad \forall k.
\end{equation}
Clearly, if $\M_i(A)=\lambda A$ with normalized $A$, then $\lambda=\bbra A|\M_i|A\kket$ and $\Re\lambda=\bbra A|\M_i^\tH|A\kket\in[\mu_{\min},\mu_{\max}]$. The quality of the bound depends on the chosen basis, i.e., the transformation $\G$ in Proposition~\ref{sttmnt:BT-nH}, and one can try to apply non-orthogonal transformations in the blocks to decrease $\mu_{\max}$. In the best case, the Hermitian and anti-Hermitian components of $\M_i$ commute such that the $\M_i^\tH$ eigenvalues coincide with the real parts of the $\M_i$ eigenvalues.

\emph{Gershgorin circle theorem.} -- A well-known theorem for bounding the eigenvalues of matrices is Gershgorin's circle theorem \cite{Gerschgorin1931-7,Golub1996,Horn2012}. It states that each eigenvalue of a square $n\times n$ matrix $\M$ is contained in one of $n$ disks in the complex plane, centered at the diagonal elements $\M_{i,i}$ with radii $R_i=\sum_{j\neq i}|\M_{i,j}|$, i.e.,
\begin{equation}
	\lambda_k(\M)\in \bigcup_i \{z\in\mathbb{C}\,\big|\, |z-\M_{i,i}|\leq R_i\}.
\end{equation}
Alternatively, one can choose the radii as the column sums instead of the row sums such that $R_i=\sum_{j\neq i}|\M_{j,i}|$. Again, the quality of the resulting bound on the largest eigenvalue real part depends on the chosen transformation $\G$. In our case, the circle theorem is only useful for diagonally dominant matrices.

All three approaches allow us to assess the gap of Liouvillians that can be transformed to a block-triangular form \eqref{eq:BTH} through properties of the blocks $\M_i$ on the diagonal instead of the entire Liouvillian. The issue may then reduce to a simple few-particle problem, allow for the application of techniques for integrable systems, or substantially reduce the cost for numerical methods. One can apply exact diagonalization for individual blocks or employ variational techniques like tensor network states \cite{White1992-11,Schollwoeck2011-326,Niggemann1997-104,Verstraete2004-7,Vidal-2005-12,Evenbly2009-79,Barthel2009-80,Corboz2009-80}.
Deriving in such a way lower bounds on the spectral gap, one can exclude dissipative phase transitions.

\section{Spectral bounds for spin-1/2 systems with spontaneous emission}\label{sec:BT-nH-example}
Let us illustrate how the spectral gap can be bounded by bringing the Liouvillian into a block-triangular form as in Proposition~\ref{sttmnt:BT-nH} and by then assessing the largest eigenvalues of the Hermitian components of the blocks $\M_i$ on the diagonal; cf.\ Eq.~\eqref{eq:Bound-H}. In particular, we consider some simple Markovian lattice models $\L=\H+\gamma\D^e$ of two-level systems (spins-1/2, qubits) with various Hamiltonians and spontaneous emission $\D^e=\sum_i\D^e_i$ corresponding to Lindblad operators $L_i=\s^-_i$, i.e.,
\begin{equation}\label{eq:L-H-De}
	\L(\dm)=-i[H,\dm]+ \gamma\sum_i \Big(\s_i^- \dm \s_i^+-\frac{1}{2} \{P^\ua,\dm\}\Big).
\end{equation}
Such models are, for example, naturally realized in Rydberg atom experiments \cite{Jaksch2000-85,Lukin2001-87,Weimer2010-6}, where $\s_i^-$ corresponds to a spontaneous deexcitation of atom $i$ that is accompanied by photon emission.
In the following, we set $\gamma=1$.

\subsection{Magnetic fields}\label{sec:De-magnField}
If the Hamiltonian only contains a magnetic field term $H^\alpha:=-\sum_i h^\alpha_i\sigma_i^\alpha$, the issue reduces to a single-site problem. In the Pauli operator basis
\begin{equation}\label{eq:B-Pauli}
	\B:=(\id,\s^z,\s^x,\s^y),
\end{equation}
the dissipator takes the block-triagonal form
\begin{equation}\label{eq:De-B}
	[\D^e_i]_\B=\begin{pmatrix}0&0&0&0\\-1&-1&0&0\\0&0&-\frac{1}{2}&0\\0&0&0&-\frac{1}{2}\end{pmatrix}.
\end{equation}
The single-site Liouvillians $\H^z_i+\D^e_i$ and $\H^y_i+\D^e_i$ for fields in the $z$ and $y$ directions are in basis $\B$ given by
\begin{equation*}
	\begin{pmatrix}0&0&0&0\\-1&-1&0&0\\0&0&-\frac{1}{2}&2ih^z_i\\0&0&2ih^z_i&-\frac{1}{2}\end{pmatrix}\quad \text{and}\quad
	\begin{pmatrix}0&0&0&0\\-1&-1&2ih^y_i&0\\0&2ih^y_i&-\frac{1}{2}&0\\0&0&0&-\frac{1}{2}\end{pmatrix},
\end{equation*}
respectively. They are lower block-triangular, but the occurring $2\times 2$ blocks are neither Hermitian nor anti-Hermitian. 

In the case of the $z$ field, the eigenvalues of the bottom right $2\times 2$ block are $-1/2\pm 2ih^z_i$. Their real part agrees with the eigenvalues of the Hermitian component \eqref{eq:MH}, which is simply $\diag(-1/2,-1/2)$.

In the case of the $y$ field, the eigenvalues of the central $2\times 2$ block are
\begin{equation}\label{eq:De-hy}
	\lambda_\pm=-3/4\pm \sqrt{1/4^2-(2 h^y_i)^2}.
\end{equation}
For $|h^y_i|\geq 1/8$, they are complex with real part $-3/4$. For $|h^y_i|<1/8$, the eigenvalues are real, going to $\lambda_\pm=-1/2,-1$ in the limit $h^y_i\to 0$. The latter two values agree with the spectral bounds set by the Hermitian component $\diag(-1,-1/2)$. In fact, for any field in the $xy$ plane, the eigenvalues of the single-site Liouvillian are given by the expressions in Eq.~\eqref{eq:De-hy} in addition to $\lambda=0,-1$. This is due to the $z$ rotation symmetry of $\D^e$.

For a magnetic field in a generic direction with $H=-\sum_{i,\alpha} h^\alpha_i\sigma_i^\alpha$, the bottom right $3\times 3$ block in the single-site Liouvillian has no zero elements. The exact eigenvalues are lengthy analytical expressions which fall into the interval $[-1,-1/2]$ in accordance with the Hermitian component $\diag(-1,-1/2,-1/2)$.

Thus, in all cases, the gap is bounded by $\Delta\geq 1/2$.

\subsection{XX and YY interactions with field}
Next, let us discuss systems with XX interactions and fields in $x$ direction on more or less arbitrary lattices, i.e., systems with the dissipator $\D^e$ as in Eq.~\eqref{eq:L-H-De} and Hamiltonians of the form
\begin{equation}\textstyle
	H=\sum_{i,j}J_{i,j}\s^x_i\s^x_j-\sum_i h_i\s^x_i
\end{equation}
with scalar coupling coefficients $J_{i,j}$.

We again employ the Pauli single-site basis \eqref{eq:B-Pauli}, labeling the elements by $(e,z,x,y)$. The field term $-i[\s^x_i,\cdotp]$ only generates transitions $y\leftrightarrow z$ on site $i$. The XX interaction term $-i[\s^x_i\s^x_j,\cdotp]$ only generates transitions $e z\leftrightarrow xy$ and $ey\leftrightarrow xz$ (and reflections) on bond $(i,j)$. Thus, the number $N_y+N_z$ of $y$ and $z$ sites is conserved under the Hamiltonian. This implies that the Liouvillian is lower block-triangular if we choose the identity as the first basis element and order all further basis elements by increasing $N_y+N_z$.

The first block $\M_0$ on the diagonal is $1\times 1$, giving the zero eigenvalue of the steady state. The basis for the second block $\M_1$ on the diagonal consists of all product operators with zero $y$ and $z$ sites ($N_y+N_z=0$) and $N_x\geq 1$ $x$ sites. As $\M_1$ does not contain any Hamiltonian matrix elements, it is diagonal, giving eigenvalues $-N_x/2$. Here, each $x$ site contributes with $-1/2$ according to Eq.~\eqref{eq:De-B}.

The basis for the third block $\M_2$ consists of all product operators that are dynamically connected to the operator with $N_y+N_z=1$ and $N_x=0$. The Hamiltonian cannot move the $y/z$ excitation, but sites that are connected to the $y/z$ site through the XX interaction can oscillate between $e$ and $x$ as discussed above. The Hermitian component $\M_2^\tH$ only contains $\D^e$ terms and is hence diagonal. The largest $\M_2^\tH$ eigenvalue is $-1/2$, corresponding to a single $y$ site. With the lattice coordination number $\zeta$, the smallest $\M_2^\tH$ eigenvalue is $-1-\zeta/2$, corresponding to one $z$ site and $\zeta$ connected $x$ sites. For small $\zeta$, the exact spectrum of the blocks with $N_y+N_z=1$ can be determined analytically.

The blocks with $N_y+N_z>1$ contain multiple $y/z$ sites. These are still immobile but can now interact via their clouds of surrounding $x$ sites if the ``distance'' is less than twice the interaction range. The Hermitian components \eqref{eq:MH} of the block matrices are again diagonal, with the largest eigenvalue being $-(N_y+N_z)/2$.

In conclusion, the largest eigenvalue real part for every Liouvillian block $\M_{n>0}$ is bounded from above by $-1/2$. Thus, irrespective of lattice geometry and (in)homogeneity of couplings $J_{i,j}$ and fields $h_i$, the gap is always $\Delta=1/2$, corresponding to a single $x$ site. Due to the symmetry of $\D^e$, this result readily generalizes to interactions and fields that are rotated by any angle around the $z$ axis, such as YY interactions and $y$ fields.

\subsection{Magnetization-conserving Hamiltonians}
If the Hamiltonian conserves the total $z$ magnetization, $[H,\sum_i\s^z_i]=0$, we can employ the single-site operator basis
\begin{equation}\label{eq:Bz}
	\B_z:=(|\da\ket\bra\da|,\,|\ua\ket\bra\ua|,\,|\ua\ket\bra\da|,\,|\da\ket\bra\ua|).
\end{equation}
In this basis, the dissipator is upper block-triangular with
\begin{equation}\label{eq:De-Bz}
	[\D^e_i]_{\B_z}=\begin{pmatrix}0&1&0&0\\0&-1&0&0\\0&0&-\frac{1}{2}&0\\0&0&0&-\frac{1}{2}\end{pmatrix}.
\end{equation}
The Hamiltonian term $\H=-i[H,\cdotp]$ conserves the numbers of up sins $N_k$ and $N_b$ in the ket and bra spaces, where the elements of the site basis \eqref{eq:Bz} have quantum numbers $(N_k,N_b)=(0,0),(1,1),(1,0),(0,1)$. As the off-diagonal term in $[\D^e_i]_{\B_z}$ simultaneously decreases both quantum numbers, the full Liouvillian \eqref{eq:L-H-De}, assumes upper block-triangular form if we order the operator basis by increasing $N_k+N_b$. For fixed $N_k+N_b$, we can for example order by increasing $N_k$. The block matrices on the diagonal are
\begin{equation}
	-(N_k+N_b)/2+\H|_{N_k,N_b},
\end{equation}
where the second term comprises the Hamiltonian matrix elements with $N_k$ and $N_b$ up spins. The excitations (block-matrix eigenvectors) are operators $|E_k\ket\bra E_b|$ with $H$ eigenstates $|E_k\ket$ and $|E_b\ket$ from the $N_k$ and $N_b$ up-spin subspaces. Their eigenvalues are
\begin{equation}
	-(N_k+N_b)/2-i(E_k-E_b).
\end{equation}
Thus, the spectral gap is again $\Delta=1/2$, irrespective of the detailed Hamiltonian.
This simple scenario, of a Hamiltonian with a conserved charge and charge-decreasing Lindblad operators can also be used to construct exactly solvable open quantum systems \cite{Torres2014-89,Nakagawa2021-126}.

\section{Spectral bounds for quadratic fermion and boson systems}\label{sec:BT-nH-quadratic}
References~\cite{Barthel2021_12,Zhang2022-129} discuss another class of open many-body systems for which $\L$ can be brought into a block-triangular form. Consider a Markovian system of identical fermions or bosons with annihilation and creation operators $a_j$ and $a^\dag_j$ that obey the canonical (anti-)commutation relations. If the Hamiltonian is bilinear in the ladder operators and Lindblad operators are either linear or bilinear and Hermitian in the ladder operators, we call such a system \emph{quadratic}. In circuit QED systems \cite{Hartmann2006-2,Hartmann2016-18,Fitzpatrick2017-7}, for example, linear Lindblad operators arise naturally from photon loss and pump process, while the coupling of cavities can lead to bilinear Lindblad operators \cite{Marcos2012-14,Tomadin2012-86}.

As detailed in Ref.~\cite{Barthel2021_12}, one can always express the Liouville super-operator of a quadratic system in terms of a suitable set of ladder super-operators \cite{Schmutz1978-30,Harbola2008-465,Prosen2008-10,Prosen2010-43,Barthel2021_12} such that it assumes a block-triangular form in the corresponding biorthogonal operator Fock basis. One can efficiently determine or bound the spectra of the smallest blocks, e.g., to establish that a model is gapless. Examples for gapless fermionic and bosonic systems are given in Ref.~\cite{Zhang2022-129}. Conversely, one can exploit the block-triangular structure to establish that the dissipative gap is at least $\kappa$ if we add linear Lindblad operators $L_j=\sqrt{2\kappa}\,a_j$ for bosons or
\begin{equation}\textstyle
	L_{j+}= \sqrt{\frac{\kappa}{2}}\,(a_j+a_j^\dag)\ \ \text{and}\ \
	L_{j-}=i\sqrt{\frac{\kappa}{2}}\,(a_j-a_j^\dag)
\end{equation}
for fermions. This implies, for example, that, without symmetry constraints beyond invariance under single-particle basis transformations and fermionic particle-hole symmetry, all gapped quadratic systems belong to the same phase \cite{Zhang2022-129}.

\section{Discussion}
In Sec.~\ref{sec:Z2model}, we discussed a non-trivial spin-1/2 model $\L=\gamma_x \D^x+\gamma_f(\D^++\D^-)+\gamma_z\D^z$, where dissipator $\D^x$ tries to polarize all spins in the $+x$ direction and the $\D^\pm$ dissipators stabilize the all-up and all-down states. The competition between the different terms leads to $\mathbb{Z}_2$ symmetry breaking at the ferromagnetic point $\gamma_x=0$. Perturbation theory suggested that the Liouville gap is $\Delta=\gamma_x/2$, independent of the other system parameters. In Sec.~\ref{sec:Real}, we introduced the general class of Liouvillians that, after an appropriate transformation, become block-triangular with Hermitian blocks on the diagonal (BTH). This implies that all eigenvalues are real and that eigenvalues obey Weyl ordering relations. A special case are Davies generators, which arise in the context of thermalization; they have full-rank steady states and are actually entirely Hermitian after an appropriate basis transformation. The model of Sec.~\ref{sec:Z2model} is an example that does not fall in the Davies class and has a non-trivial block-triangular structure. A very simple non-Davies example is a spin-1/2 with Lindblad operator $\s^-$ and pure steady state $|\da\ket\bra\da|$. When considering the competition between several BTH Liouvillian terms, the Weyl ordering relations imply that the gap of each individual term is a lower bound for the gap of the full Liouvillian. For a large class of models, this excludes phase transitions. On the other hand, for local couplings, the nonzero gap enables an efficient preparation of the corresponding steady states \cite{Barthel2012-108b} and makes them stable with respect to perturbations \cite{Cubitt2015-337}.

In Sec.~\ref{sec:BT-nH}, we considered Liouvillians that can be transformed into a block-triangular form but have non-Hermitian blocks on the diagonal. There are several paths to bound the Liouvillian gaps of such systems, e.g., through the block singular values, Gershgorin's circle theorem, or, maybe most effectively, through the Hermitian components of the block matrices. In Sec.~\ref{sec:BT-nH-example}, we saw how the latter approach can be used to establish nonzero gaps for classes of lattice spin models with spontaneous emission, e.g., for Hamiltonians with magnetic fields and XX or YY interactions with almost no restrictions on lattice geometry, interaction range, and spatial variations.

For now, our methods for finding operator transformations that block-triangularize Liouvillians were based to the use of symmetries and further dynamical constraints as demonstrated in the examples of Secs.~\ref{sec:Z2model}, \ref{sec:Real-Z2model}, \ref{sec:BT-nH-example}, and \ref{sec:BT-nH-quadratic}. It would be great to develop more sophisticated approaches.

Of course, there exist models, where the discussed conditions that exclude dissipative phase transitions are not met. A classic example is the phase transition in the driven-dissipative Kerr oscillator \cite{Drummond1980-13,Bartolo2016-94,Zhang2021-103,Rodriguez2017-118}. Here, the block-triangular structure of quadratic models (Sec.~\ref{sec:BT-nH-quadratic}) is spoiled by the quartic Hamiltonian term $a^\dag a^\dag a a$, which is due to photon-photon interaction. A similar interaction term spoils the block-triangular structure in the driven-dissipative Bose-Hubbard model, and a steady-state phase transition is expected to occur in two dimensions \cite{FossFeig2017-95,Vicentini2018-97}.

As seen in the examples, the block-triangular structures may imply that perturbative treatments actually yield exact Liouvillian eigenvalues. In general, reducing the spectral problem to the blocks on the diagonal substantially decreases the costs for numerical approaches like exact diagonalization and can make variational techniques like tensor network state methods \cite{White1992-11,Schollwoeck2011-326,Niggemann1997-104,Verstraete2004-7,Vidal-2005-12,Evenbly2009-79,Barthel2009-80,Corboz2009-80} applicable. It is also conceivable that the block structure may, in some models, allow for the application of methods for integrable systems like the Bethe ansatz \cite{Bethe1931,Korepin1993}, Luttinger liquid theory \cite{Luttinger1963-4,Giamarchi2004}, or conformal field theory \cite{Belavin1984-241,Francesco1997}.

\begin{acknowledgments}
We gratefully acknowledge helpful discussions with Jianfeng Lu, Hendrik Weimer, and Xin Zhang, and
support through U.S.\ Department of Energy grant DE-SC0019449.
\end{acknowledgments}

\end{document}